\documentclass[12pt]{article}
\usepackage{amssymb}
\usepackage[pdftex]{graphicx}

\title{Algorithms for Proportional Representation in Parliament in Divisor and Multiplicative Form}
\author{Ra\'ul Rojas\\ Department of Mathematics and Statistics\\University of Nevada Reno}
\begin{document}
\maketitle

\begin{abstract}
We consider three algorithms for allocating parliamentary seats by proportional representation. The usual approach  to describing such algorithms is to compute a quota of votes that each party uses
  to ``acquire'' representatives. This kind of description follows a divisor method, since the number of representatives for a party is equal to the number of votes for that party, divided by the quota. We show that a simple multiplicative form with different rounding methods produces algorithms equivalent to the divisor  methods. The multiplicative form is  intuitive and easier to understand for a wider audience.
\end{abstract}

\section{Proportional representation}

The question of apportioning seats in parliament according to the number of votes received by the parties involved has been discussed since there have been parliaments. Thomas Jefferson famously described such a method in a letter to George Washington in 1792. The algorithm had also been invented by the Belgian mathematician d'Hondt in
1878. Many other algorithms and alternatives were developed in the 20th century, especially in Europe. The usual description of such methods is to find the number of
votes needed to ``acquire'' a representative. For example, if three parties received 600, 300, and 100 votes, respectively, a quota of 100 votes for a representative  
would give each party 6, 3, and 1 representatives, respectively. If there are only 10 seats, each party's representation is exactly proportional to its vote total. In this simple case, we can simply divide the number of votes for each party by the quota. However, if the quota is not an integer, the division will generally produce fractional results, that is, an integer plus a remainder less than one. In this
case, we have to decide how to allocate the last available seats to the different parties.

In this paper we look at three algorithms and the way they allocate seats. We show that while formulating the algorithms in terms of quotas is the traditional way of describing them, a simple multiplicative form is more intuitive and helps to understand exactly how the algorithms differ, and what kind of advantage they give to smaller or larger parties.
We examine the Hare-Niemeyer algorithm, the Jefferson-d'Hondt method, and the Sainte-Lagu\"{e} algorithm. Descriptions of these and many other methods
can be found in extensive handbooks on the mathematics of elections \cite{wallis, pukel}. A general introduction to election math can be found in \cite{sullivan}. Here we want to simplify the exposition and make the general intuition of proportional representation methods understandable to a wider audience.

Consider elections to a parliament where $N$ seats are to be distributed among $k$ parties according to their share of the vote (proportional representation). We have the following situation:
\begin{itemize}
\item Each party $i$ receives $v_i$ votes, for $i=1,\ldots,k$. 
\item The fraction of votes for the $i$-th party is $f_i=v_i/V$, where $V$ is the total number of votes for all parties. 
\item Each party $i$ is allocated $n_i$ seats. We want to achieve  $n_1+\ldots +n_k=N$, possibly iteratively.
\item The
lower quota of seats for the $i$-th party is defined as the integer $l_i= \lfloor N f_i \rfloor$, while $u_i= \lceil N f_i \rceil$ is the upper quota of seats.
\item The residual for each party is the
number $r_i=N f_i - n_i$.
\end{itemize}

When we assign $n_i$ seats to each party $i$, for $i=1,\ldots,k$, we want  the ratio $n_i/N$ to be as close to $f_i$ as possible. 
Depending on how we want to measure ``fairness'', different algorithms can be used. 

Divisor algorithms work under the theoretical assumption
that parties ``acquire'' seats with votes. The number of votes needed to win a seat is called the ``quota''. In the case that $N$ seats in parliament
correspond to $V$ votes, the ideal quota is $q=V/N$. If we divide the number of votes of each party by this quota, we get 
$$
n_i = \frac{v_i}{q}=N\frac{v_i}{V}=Nf_i
$$
The problem with this simple computation is that $q$ and $n_i$ can be non-integers. All divisor algorithms for
proportional representation have to produce $n_i$ as an integer, and most of them compute first a quota $q$. Let us consider some ways
for achieving proportional representation.

\subsection{The Hare-Niemeyer algorithm}

A simple seat assignment algorithm is the following:
\begin{itemize}
\item Assign each party $i$ its lower quota of seats $n_i=l_i= \lfloor N f_i \rfloor$.
\item If the total number of seats $N$  has already been distributed, the algorithm stops.
\item If it has not, it continues to allocate seats to parties in the higher to lower order of their (positive) residuals until all representatives have been allocated.
 If two residuals are equal, we select one of them at random.
\end{itemize}
Note that we can implement randomness of selection by adding a different small random number $\epsilon$ to equal residuals, so that they are made different.

This
algorithm satisfies the so-called ``quota property'' because for the $n_i$ representatives assigned to the $i$-th party, it holds that $l_i\leq n_i \leq u_i$, by construction.  No party is assigned representatives below its lower quota of seats, nor above its upper quota.
Another way of saying the same thing is  that no party can complain of having a deficit of seats (relative to the ideal number $Nf_i$) greater than one, and also no party obtains an
excess of seats higher than one. In a parliament with 500 seats and ten parties, the upper bound for the absolute sum of deficit/excess of seats for all parties is less than 10, which seems acceptable.

Note that the Hare-Niemeyer algorithm in multiplicative form can be written in divisor form, by setting $q=V/N$ and 
computing the provisional assignment of seats $n_i=\lfloor v_i/q \rfloor$. If not all $N$  seats have yet been assigned, we assign 
the missing seats to the parties in descending order of their residuals $v_i/q - \lfloor v_i/q \rfloor$.

\subsection{The d'Hondt-Jefferson algorithm}

The d'Hondt-Jefferson algorithm (dHJ in what follows) is usually explained in terms of an adaptive quota $q$ (the votes ``paid'' for a seat), which is adjusted
iteratively until all seats have been allocated. Note, however, that any algorithm that proposes a quota $q$ and a number of seats
for each party $i$ equal to some kind of integer rounding of $v_i/q$, can be transformed into an algorithm that computes the same rounding for  $Qv_i$, where $Q$ is the reciprocal of $q$.

The classical dHJ algorithm proceeds as follows: we start with a table of votes and allocated seats $n_i$ for each party, which are all zero at the beginning.
$$
   \begin{array}{|c|c|c|c|c|} \hline
 \rm{seats}&     0 & \ldots & 0 & \ldots \\ \hline
   \rm{present \ quota} &         \Box & \ldots & \Box & \ldots \\ \hline
   \rm{next \ quota} &    v_1/1 & \ldots & v_j/1 & \ldots \\ \hline
   \end{array}
$$
Without loss of generality, assume that the first party has a majority of votes and gets assigned the first seat. The table is updated to
show how many votes are  needed for one seat, if nothing else happens, and how many votes would be needed for each seat, if that party is
assigned a second seat in a future step (``next quota'' row). For  parties with no change, we leave their column unchanged.
$$
   \begin{array}{|c|c|c|c|c|} \hline
 \rm{seats}&           1 & \ldots & 0 & \ldots \\ \hline
   \rm{present \ quota} &      v_1/1 & \ldots & \Box & \ldots \\ \hline
     \rm{next \ quota} &           v_1/2 & \ldots & v_j/1 & \ldots \\ \hline
   \end{array}
$$
We only need to keep the last two rows of the algorithm's history. After $N$ iterations we have the situation shown below, where the first party
has  been allocated $n_1$ seats and the $j$-th party has $n_j$  seats.
$$
   \begin{array}{|c|c|c|c|c|} \hline
  \rm{seats}&     n_1 & \ldots & n_j & \ldots \\ \hline
    \rm{present \ quota} &     v_1/n_1 & \ldots & v_j/n_j & \ldots \\ \hline
     \rm{next \ quota} &            v_1/(n_1+1) & \ldots & v_j/(n_j+1) & \ldots \\ \hline
   \end{array}
$$
In each step of the algorithm we choose the maximum number in the last row (i.e., the highest bid of votes that a party can pay on average for its representatives if it is assigned one more). Let us assume that the last maximum was found in column $j$. The number $v_j/n_j$
represents the average number of votes that party $j$  ``pays'' for each of its seats. We are assuming that we have already distributed $N$ seats across all parties, and that  the last seat was  allocated
to party $j$. We stop the algorithm, and $q=v_j/n_j$ is called the quota
of votes paid in the last step for a seat. We assume for simplicity that all the numbers in the last row of the table were different (if not, we add a small random number to those that are the same).

Now consider what happens to the seats allocated to party 1 (which could be any  party other than $j$). Since $v_1/(n_1+1)<v_j/n_j$ (because we picked column $j$ in the last step) 
and $v_1/n_1>v_j/n_j$, because the last seat of party 1 was allocated before the last seat of party $j$. Let us define $q=v_j/n_j$. If we divide the votes of party 1 by  $q$, we get
$$
\frac{v_1}{v_1/n_1} < \frac{v_1}{q} < \frac{v_1}{v_1/(n_1+1)}
$$
that is
$$
n_1 < \frac{v_1}{q} < n_1+1
$$
If we take the integer part of $v_1/q$ we obtain
$$
\left \lfloor{\frac{v_1}{q}}\right \rfloor = n_1
$$
But the same result can be obtained by defining $M=V/q$, where $V$ is the total number of votes. In this case
$$
 \lfloor Mf_1 \rfloor =  \left \lfloor \frac{V}{q} \frac{v_1}{V} \right \rfloor = n_1 
$$
The above analysis for party 1 could be done for any other party $i$.
In the classical algorithm $q$ decreases iteratively until all seats are filled. In the multiplicative algorithm $M$ grows steadily (starting from $M=1$) until all seats are
filled, that is, for party $i$ we set $n_i=\lfloor Mf_i \rfloor$.
Thus the dHJ algorithm can be expressed as a multiplicative allocation algorithm, just as in the case of Hare-Niemeyer, and the difference will be the way of handling residuals after the initial allocation of lower quotas.
The dHJ algorithm  can be formulated using the following steps:
\begin{itemize}
\item Start with  $M$  seats to distribute among parties. We can start with $M=N$.
\item Assign each party $i$ an integer number of representatives $n_i=\lfloor M f_i \rfloor$.
\item If the total number of $N$ representatives has already been distributed, the algorithm stops. 
\item If not, $M$ is slightly increased  and we repeat the computation to assign the $n_i$ representatives, for $i=1,\ldots,k$  ($M$ does not have to be an integer). If
the algorithm overshoots and more than $N$ seats are allocated, we decrease $M$ slightly.
\end{itemize}
Note that the algorithm must stop at some point, because the sum of all $n_i$'s will eventually exceed $N$ (since we are always increasing $M$).
Note also, that we increase $M$ systematically, looking for the value where $\sum{n_i}=N$. If $\sum{n_i}$ becomes greater than $N$, we decrease $M$.
Note that if we have two or more parties with the same share of the vote, increasing $M$ may cause them both to get an extra seat in the
same iteration. This means that $\sum{n_i}$ may overshoot $N$ by one or more. In this case we ``de-assign" seats to these parties, randomly, until
the sum of allocated seats does not exceed  $N$. Alternatively, we can make the fractions of votes (the $f_i$) slightly different from the beginning by adding a small random number.

The only difference between the dHJ algorithm and the Hare-Niemeyer method is how they handle  residuals. When the parliament is large and the
number $k$ of competing parties is small, they tend to get solutions that are very similar.

\subsection{Comparing Hare-Niemeyer and d'Hondt-Jefferson}

We can better understand the difference between the dHJ algorithm and the Hare-Niemeyer method if we look at the residuals, assuming that each
party was  allocated its lower quota. Fig.1 shows the residuals for parties A,B, C, and D. Each of them has a frequency of votes $f_A,f_B,f_C,f_D$, respectively. Assume that the parties C and D obtained a higher share of the total vote than
the parties A and B ($f_C=f_D$, $f_A=f_B$, and $f_A<f_C$). If the four residuals were zero, for Hare-Niemeyer, we would be finished. The same is true for the dHJ method.

Given the  distribution of residuals shown, and assuming that we still need to allocate two more seats, in order to reach the final number $N$,
the Hare-Niemeyer method would assign them to parties A and B, because their residuals are larger. However, the dHJ method would increase the number
$M$, let's say to $M+1$. The number of seats with $M$, for party $i$, is $\lfloor Mf_i \rfloor$, but with $M+1$, it is $\lfloor Mf_i + f_i \rfloor$. That is, those parties with
larger $f_i$ have a greater chance of ``jumping'' to the next integer. This is shown in Fig. 1 . Now, parties C and D can increase their residuals above one, and they both get  the last two representatives.

As this simple example shows, Hare-Niemeyer is agnostic about how the residuals were generated and how the parties jump to the next integer.
The dHJ algorithm favors those parties with higher $f_i$, that is, the larger parties according to their share of votes. If we assume that the residuals are larger for
smaller parties (because more of their votes remain ``unused'' after a quota is set), then Hare-Niemeyer favors smaller parties.

\begin{figure}[htb]
\centerline{\includegraphics[width=12cm]{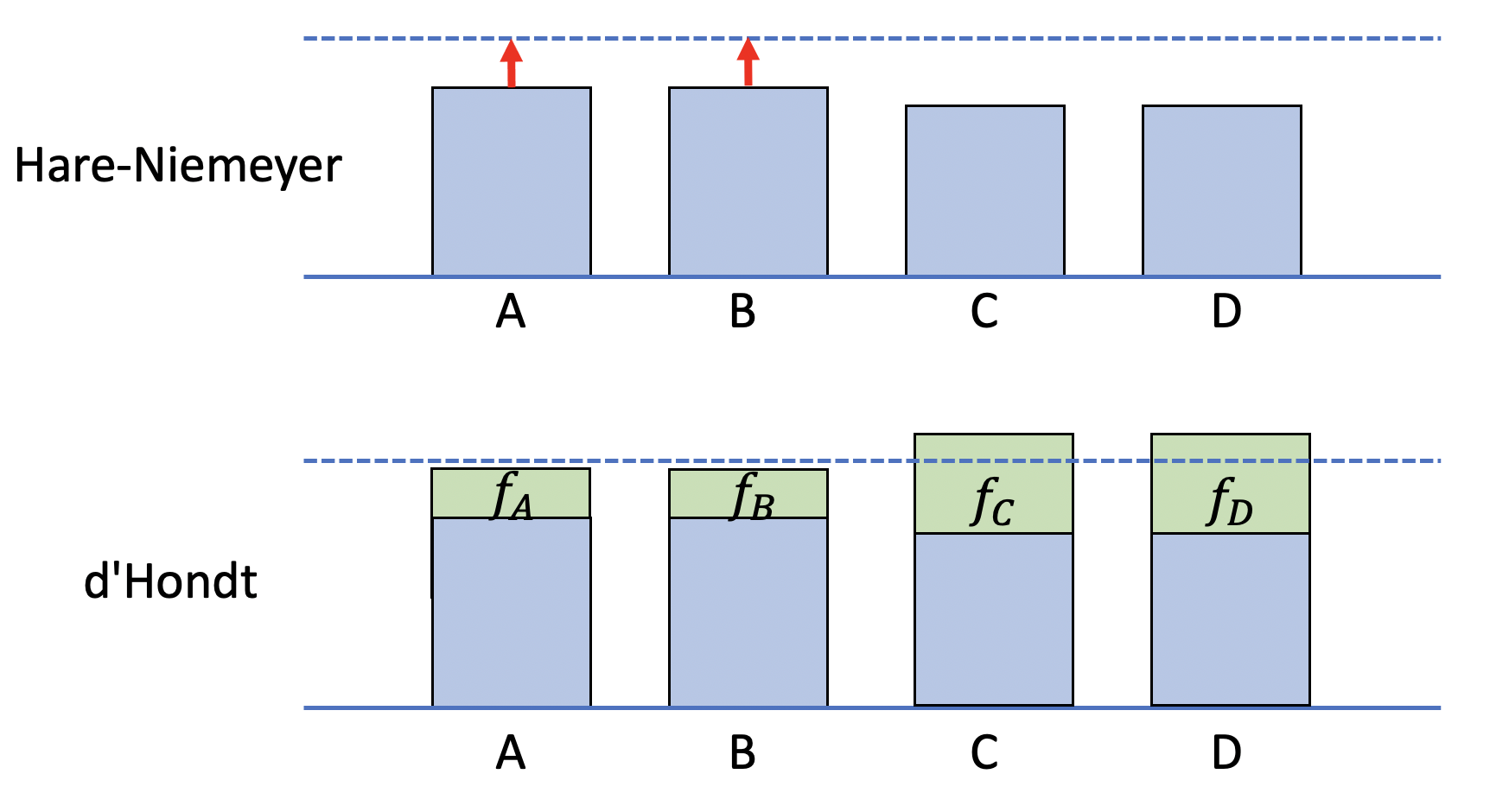}}
\caption{Jumping to the next integer in Hare-Niemeyer and in the dHJ algorithm. The residuals for each party are shown in blue. We assume that the parties C and D obtained a higher share of the total vote than
the parties A and B ($f_A=f_B$, $f_C=f_D$,  and $f_A<f_C$)\label{fig1}}
\end{figure} 

\subsection{Sainte-Lagu\"{e}-Algorithm}

There is a proportional representation method that is very similar to the dHJ method, but  differs in the way it handles residuals (when explained in the multiplicative form).
In the Sainte-Lagu\"{e} algorithm we take a number $M$ and we compute $Mf_i$ ``ideal seats'' for party $i$. This number is not truncated as in the dHJ method;
it is rounded to the nearest integer (7.6 is rounded to 8, 7.3 is rounded to 7). If the sum of the allocated seats is too large, we choose a lower $M$ for the
computation. If the sum of the allocated seats is too small (at any point in the algorithm), we choose a higher $M$. With computers, it is very easy
to slowly increase or decrease the number $M$ until the number of allocated seats  equals the number of seats to be filled. Looking back at
Fig.1, we would have a situation where all four parties are rounded to the upper quota (because their residuals are greater than 0.5). If the total exceeds the number of available seats, the number $M$ has to be
reduced until the residuals of parties C and D are rounded down. If at some point more than one party get seats  and the total exceeds the available number of seats, and those parties have identical $f_i$ values, then we have to pick one of those parties at random and de-allocate one seat. Or, as we have pointed out before, we can make the $f_i$'s different before we start, by adding a small random number that ``precomputes'' a preference for
one party over another with the same $f_i$.

Explained in the divisor form, the classical Sainte-Lagu\"{e} algorithm works exactly like the dHJ algorithm, when looking for the next quota for acquiring a seat, but the divisors
in the table are different. For $n_i$ assigned representatives, the divisor is $2n_i+1$. Let us assume that we have already assigned $N-1$ seats and that the table looks like this:
$$
   \begin{array}{|c|c|c|c|c|} \hline
  \rm{seats}&       n_1 & \ldots & n_j & \ldots \\ \hline
     \rm{present \ quota} &     v_1/(2(n_1-1)+1) & \ldots & v_j/(2(n_j-1)+1)& \ldots \\ \hline
      \rm{next \ quota} &             v_1/(2n_1+1) & \ldots & v_j/(2n_j+1) & \ldots \\ \hline
   \end{array}
$$
Assume that the maximum in the ``next quota'' row is found at column $j$. Party $j$ gets assigned the last seat. In that case, with $q=v_j/(2n_j+1)$ and by a similar argument as with the HJ-algorithm:
$$
\frac{v_1}{v_1/(2n_1-1)} < \frac{v_1}{q} < \frac{v_1}{v_1/(2n_1+1)}
$$
that is
$$
2n_1 - 1< \frac{v_1}{q} < 2n_1+1
$$

Because of the way the divisors are defined, the quota $q$ is around twice as small as with the dHJ algorithm. It makes sense
to compute $ {v_1}/{2q} $ as an approximation to the number of seats that have to be allocated to party 1,  then
$$
n_1- \frac{1}{2} < \frac{v_1}{2q} < n_1+\frac{1}{2} 
$$
Adding 1/2 to each term
$$
n_1 < \frac{v_1}{2q} +\frac{1}{2}< n_1+1
$$
and therefore 
$$
\biggl\lfloor{\frac{v_1}{2q} +\frac{1}{2}}\biggr\rfloor =n_1.
$$
If we define $M=V/(2q)$ and with $f_1=v_1/V$, we can rewrite this as
$$
\biggl\lfloor Mf_1 + \frac{1}{2}\biggr\rfloor = n_1.
$$
Here, instead of just truncating $Mf_1$ we are rounding to the nearest integer. This is because given an $x$ we obtain rounding to the nearest integer by computing $ \left \lfloor x+  \frac{1}{2}\right \rfloor $.
This shows that the multiplicative and the divisor form of the Sainte-Lagu\"{e} algorithm are equivalent, when the rounding
in the multiplicative form is done to the nearest integer.

There are some variations of the Sainte-Lagu\"{e} algorithm that differ only in the threshold for rounding up or down the fractional
values obtained from each product $Mf_i$.

\section{Seeded Algorithms}

There are cases where elections are held in two stages: for example, $D$ representatives are elected in districts and then $N$ additional
representatives are added in order to correct the distortions of the first election and to approximate the allocation of $D+N$ seats to proportional representation, according to the number of votes per party.

In this case, each party $i$ has $d_i$ representatives from the first phase of the election and receives $n_i$ additional seats (which may be zero
in some cases).

\subsection{The dHJ algorithm and Sainte-Lagu\"{e}}

In an election in two stages, we can use Hare-Niemeyer, the dHJ algorithm or Sainte-Lagu\"{e} to adapt to proportional representation. Assume that there are $D$ districts where the representatives are
elected by majority of votes, and additional representatives that are allocated in order to approximate proportional representation. The algorithm stops when a fixed number $N$ of additional
representatives has been allocated, or, alternatively, when the residual of each party is less than one in absolute value.

First, we can consider dHJ and Saint-Lagu\"{e}. We start with $M=D+1$ and compute $Mf_i$ for each party. Then
\begin{itemize}
\item If $d_i \geq Mf_i$,  we set $n_i=0$,
\item If $d_i \leq Mf_i$, we set $n_i= R(Mf_i-d_i)$, where $R$ is a rounding function. For dHJ we round down all residuals, for Saint-Lagu\"{e}
we round to the nearest integer.
\item If the stop criterion has been reached, we stop. If not, we increase $M$ and continue. If an upper bound $N$ for the additional seats has been set, and it is exceeded, we decrease $M$
and continue.
\end{itemize}
%A similar adjustment can be made with the Hare-Niemeyer method: we set $M=D+N$ and compute $Mf_i$ for each party. Parties for which
%$d_i \geq Mf_i$ do not get any additional seats. All other parties receive $\lfloor Mf_i \rfloor - d_i$ additional seats. If the number of total seats does not surpass $D+N$ seats, we assign the missing seats to parties with a total of seats below $\lfloor Mf_i \rfloor +1$, according to the value of their residuals after truncation of $Mf_i-d_i$, from largest to smallest.

\subsection{The sequential Hare-Niemeyer algorithm}

The classical Hare-Niemeyer method can be implemented iteratively, using the following sequential algorithm: 
\begin{itemize}
\item For each party $i$, its ideal due number of representatives is computed as $Nf_i$. 
\item Each party $i$ starts with an initial allocation of representatives $n_i=0$ . 
\item The $N$ representatives to be allocated are
assigned  to the parties one by one. In the $j$-th iteration we select the party $p$ with the largest deficit of assigned representatives, that is, the largest
$Nf_i-n_i$ (if there are several that are the highest and equal, we choose one at random). Note that the deficits $Nf_i-n_i$ are real values.
\item  For the next iteration we reset $n_p:=n_p+1$.  That is, we
assign one more representative  to party $p$.
\item  We continue in this way until all the
$N$ representatives have been assigned in $N$ iterations.
\end{itemize}
The algorithm cannot assign a party more than its upper cuota, because once a party reaches its upper cuota, it no longer has a deficit of
representatives and will not participate in subsequent iterations. It must  assign each party at least its lower cuota, in some iteration. When the algorithm
is running, a party with a residual $Nf_i-n_i$ less than one will not receive a new representative if there are parties with residuals greater than one. This means that
at some point all residuals will be lower than one, by the construction of the algorithm, and this is when the algorithm concludes step 1
of the classical Hare-Niemeyer method. The final iterations of the algorithm then simply implement the second step of the Hare-Niemeyer algorithm, if necessary (i.e., in the case that not all $Nf_i$ are integers). 

\subsection{Seeded sequential Hare-Niemeyer}

The sequential variation of the Hare algorithm is important when we have an initial distribution of $D$ representatives that we want to supplement with additional representatives
in order to obtain proportional representation.

In order to approximate the final percentage of representatives for each party to its percentage of votes. We do the following:
\begin{itemize}
\item Let's assume that each party has been already assigned a number $d_i$ of representatives (in electoral districts, for example). The total number of representatives
for all districts is D.
\item We set $m_i=d_i$ for all parties, numbered from 1 to $k$.
\item  If we want  to assign
additional representatives, we compute the deficit of seats for each party, in the $j$-th iteration, as $f_i(D+j)-m_i$. A negative deficit (surplus) signals an excess of seats already allocated to this party.
\item  We then assign seats,
one by one, each time to the party with the largest deficit of seats, until at iteration $J$, every party has a residual $f_i(D+J)-m_i$ which, in absolute value, 
is less than 1. Then, the algorithm stops.
\end{itemize}

Note that when the algorithm starts, we don't know how many additional representatives $J$ will be needed, this is determined when  the algorithm stops at iteration $J$.

If we consider all parties with a shortage of representatives as one block, we  only add representatives to that block. 
Obviously, at some point, one of the parties in that block must go from having a deficit to having an excess of representatives, then another party must do so, and so on. Since only one representative is added each time, the excess of representatives for a party coming from a deficit must be strictly less than one. We continue with the parties that still have a deficit.
The algorithm stops because seeding representatives sequentially reduces the deficit of the parties with deficit and  monotonically dilutes the excess of parties that received more
representatives in the district elections than their upper cuota. If a party has an excess of representatives greater than one, the algorithm continues 
distributing representatives among the parties with deficit. There is at least a party with a deficit, if a party has a surplus of representatives, so the algorithm only
stops when the absolute value of all residuals  $f_i(D+j)-m_i$ is less than one.

If we do not allow the number of additional representatives, after district elections, to go above $\ell$, we stop the algorithm at the iteration $\mathrm{min}(\ell,J)$. If $\ell < J$,  a party could still have an excess of representatives over one when we stop, but we cannot dilute that excess further down. The algorithm has made a ``best effort'' with the available number $\ell$ of  additional representatives. A better result could only be obtained by subtracting representatives from the party with the highest excess
after the district elections, but we cannot do that.

If the number of additional representatives is fixed to $T$ from the beginning, we know what is the fraction of representatives due for each party, it is $f_i(D+T)$.
In this case, we simply run the sequential Hare algorithm, seeded with $m_i=d_i$,  with $N=D+T$. If $J>T$ for the given initial distribution, the result with $T$ additional representatives will be suboptimal (in terms of the quota property). If $T>J$ the final result may  not be as good as the one found with $J$ iterations, but it will still satisfy the quota property.

An example of the last statement could be that at iteration $J$ all parties reach a residue zero (because the numbers $(D+J)f_i$ are integers). This is
a perfect result that can be spoiled by assigning one more representative to a party selected at random. 

\subsection{Epilogue}

The difficulty with proportional representation algorithms is that they are rather obscure when expressed in the divisor form. The multiplicative form
is transparent and easier to understand for a wider audience. The advantages or disadvantages for small parties are also easy to explain.
It is important to simplify the explanation of voting algorithms so that they can be understood by the general public.

\end{document}